\documentclass[10pt]{article}

\usepackage{graphicx}
\usepackage{amsmath}
\usepackage{epsf}
\usepackage{graphpap}

\def\be{\begin{equation}}
\def\ee{\end{equation}}

\begin{document}
% \footnote{}

\begin{center}{\bf  \Large Closed-form Solutions for the Lucas-Uzawa model: Unique or Multiple}\\[2ex]
{R. Naz}\\
{ Centre for Mathematics and Statistical Sciences,
  Lahore School of Economics, Lahore, 53200, Pakistan\\

$^*$  Corresponding Author Email: drrehana@lahoreschool.edu.pk.\\

}
%
%\address{\\}
\end{center}
  \begin{abstract}
 Naz and Chaudhry \cite{naz2017} established multiple closed-form
 solutions for the basic Lucas-Uzawa model. According to Boucekkine and
Ruiz-Tamarit \cite{ruiz1} and  Chilarescu \cite{chil} unique
closed-form solutions exist for the basic Lucas-Uzawa model. We
equate expressions for variables $h(t)$ and $u(t)$.  We provide here
 condition for the unique closed-form solution and proposed an open question for evaluation of integral in closed-form. A similar analysis
 is carried out for the Lucas-Uzawa model with logarithmic utility
 preferences.

\end{abstract}

\section{Introduction} The following model is
discussed for the closed form solutions by Boucekkine and
Ruiz-Tamarit \cite{ruiz1}, Chilarescu \cite{chil} and Naz and
Chaudhry \cite{naz2017} for fairly general values of parameters.
 The representative agent's utility
function is defined as
  \be Max_{c,u} \quad \int_0^{\infty}\frac{c^{1-\sigma}-1}{1-\sigma} e^{-\rho t} ,
  \; \sigma \not=1\label{(rn1)}\ee subject to the constraints
of  physical capital and human capital:
\begin{eqnarray} \dot k(t)
= \gamma k^\beta u^{1-\beta} h^{1-\beta}-\pi k-c, \; k_0=k(0) \nonumber\\
\dot h(t) =\delta(1-u)h,\; h_0=h(0) \label{(rn2)}.\end{eqnarray}

 Recently, Bethmann \cite{log1}
developed a stylized version of the two sector Lucas-Uzawa
 model with logarithmic utility preferences and solved the model by dynamic programming technique.
Chilarescu  and Sipos \cite{log2} derived closed-form solutions for
the variables in the model proposed by Bethmann in terms of
numerically computable functions involving integrals. Chaudhry and
Naz \cite{naz2018} derived multiple closed-form solutions for this
model. The representative agent's utility function is defined as
  \be Max_{c,u} \quad \int_0^{\infty} e^{-\rho t} \ln (c) dt,
 \label{(rn1)}\ee subject to the constraints
of  physical capital and human capital: \begin{eqnarray} \dot k(t)
= A k^\alpha (u h)^{1-\alpha } -c, \; k_0=k(0), \nonumber\\
\dot h(t) =\delta (1-u)h,\; h_0=h(0), \label{(rn2)}\end{eqnarray}
where
 $\rho>0$ is the discount factor, $\alpha$ is the
elasticity of output with respect to physical capital, $A>0$ is the
level of technology in the goods sector, $\delta>0$ is the level of
technology in the education sector, $k$ is physical capital, $h$ is
human capital, $c$ is per capita consumption and $u$ is the fraction
of labor allocated to the production of physical capital.

\section{Closed-form solutions for Lucas-Uzawa model: Unique or multiple }

 The following closed-form solution derived via two first integrals $I_1$ and $I_2$ is given in
equation (3.21) on page 474 of  Naz and Chaudhry \cite{naz2017}:

\begin{eqnarray} c(t)=c_0
z_0^{\frac{\beta}{\sigma}}e^{-\frac{(\rho-\delta
)}{\sigma}t}z^{-\frac{\beta}{\sigma}}, \nonumber\\
k(t)=\bigg(\frac{k_0}{c_0z_0^{\frac{\beta-\sigma}{\sigma}}}-F(t)
\bigg) c_0z_0^{\frac{\beta}{\sigma}}z(t)^{-1}e^{\frac{(\delta
+\pi-\pi \beta)}{\beta}t}
,\nonumber\\
 h(t)=\frac{h_0}{z_0[\sigma c_0 z_0^{\beta-1}
-(\rho+\pi-\pi \sigma)k_0 z_0^{\beta-1}+\beta \gamma(1-\sigma)k_0]}[
\sigma c_0 z_0^{\frac{\beta}{\sigma}}e^{-\frac{(\rho-\delta
)}{\sigma}t}z^{-\frac{\beta}{\sigma}+\beta} \nonumber\\
+(\beta \gamma (1-\sigma) -(\rho+\pi-\pi \sigma)z^{\beta-1})
(\frac{k_0}{c_0z_0^{\frac{\beta-\sigma}{\sigma}}}-F(t) )
c_0z_0^{\frac{\beta}{\sigma}}e^{\frac{(\delta+\pi
-\pi \beta)}{\beta}t}], \nonumber\\
 u(t)=\frac{u_0}{k_0}[\sigma c_0 z_0^{\beta-1}-(\rho+\pi-\pi \sigma)k_0 z_0^{\beta-1}+\beta \gamma(1-\sigma)k_0]\nonumber\\
 \times \frac{(\frac{k_0}{c_0
z_0^{\frac{\beta-\sigma}{\sigma}}}-F(t))}{[\beta
\gamma(1-\sigma)-(\rho+\pi-\pi \sigma)
 z^{\beta-1}](\frac{k_0}{c_0 z_0^{\frac{\beta-\sigma}{\sigma}}}-F(t))+\sigma z^{\beta-\frac{\beta}{\sigma}}e^{-(\frac{\delta+\pi-\pi
\beta}{\beta}-\frac{\delta-\rho}{\sigma} )t}},\nonumber\\
 \lambda(t)=c_0^{-\sigma}z_0^{-\beta}e^{(\rho-\delta)t}z^{\beta},\nonumber\\
 \mu(t)= c_1 e^{(\rho-\delta )t}
\nonumber,\end{eqnarray} where \begin{eqnarray} F(t)=\int_0^t
z(t)^{\frac{\sigma-\beta}{\sigma}} e^{-(\frac{\delta+\pi-\pi
\beta}{\beta}-\frac{\delta-\rho}{\sigma} )t} dt,\nonumber\\ z(t)=
\frac{z^*z_0}{[(z^{*1-\beta}-z_0^{1-\beta})e^{-\frac{(1-\beta)(\delta
+\pi)}{\beta}t}+z_0^{1-\beta}]^{\frac{1}{1-\beta}} },\label{sol1}\\
\lim_{t\to\infty}
F(t)=\frac{k_0}{c_0z_0^{\frac{\beta-\sigma}{\sigma}}},\nonumber\\
 \rho<\delta <\rho+\delta \sigma, \frac{\delta+\pi-\pi
\beta}{\beta}-\frac{\delta-\rho}{\sigma}
>0,\nonumber\\
c_0 z_0^{\frac{\beta}{\sigma}}=\bigg(\frac{c_1
\delta}{(1-\beta)\gamma}\bigg)^{-\frac{1}{\sigma}},\nonumber\\
\frac{\gamma(1-\beta)(\rho-\delta+\delta \sigma)}{\delta}\nonumber\\=\frac{u_0}{k_0}[\sigma c_0 z_0^{\beta-1}-(\rho+\pi-\pi \sigma)k_0 z_0^{\beta-1}+\beta \gamma(1-\sigma)k_0],\nonumber\\
 z^*=\bigg(\frac{\beta \gamma}{\delta +\pi} \bigg
)^{\frac{1}{\beta-1}}\nonumber.\end{eqnarray}

The following closed-form solution via one first integral $I_1$is
given in equation (4.6) on page 476 of  Naz and Chaudhry
\cite{naz2017}:

\begin{eqnarray} c(t)=c_0
z_0^{\frac{\beta}{\sigma}}e^{-\frac{(\rho-\delta
)}{\sigma}t}z^{-\frac{\beta}{\sigma}}, \nonumber\\
k(t)=\bigg(\frac{k_0}{c_0z_0^{\frac{\beta-\sigma}{\sigma}}}-F(t)
\bigg) c_0z_0^{\frac{\beta}{\sigma}}z(t)^{-1}e^{\frac{(\pi+\delta
-\pi \beta)}{\beta}t}
,\nonumber\\
h(t)=\bigg[\bigg(\frac{(\delta+\pi)(1-\beta)}{\beta} \frac{k_0 }{c_0
z_0^{\frac{\beta-\sigma}{\sigma}} } +\frac{\delta u_0 k_0 }{c_0
z_0^{\frac{\beta-\sigma}{\sigma}} }-\delta u_0
G(t)\bigg)e^{ -\frac{(\delta+\pi)(1-\beta)}{\beta}t}\nonumber\\
-\delta u_0 (\frac{k_0 }{c_0 z_0^{\frac{\beta-\sigma}{\sigma}}}-
F(t))\bigg]\times
\frac{c_0z_0^{\frac{\beta}{\sigma}}}{\frac{(\delta+\pi)(1-\beta)}{\beta}
u_0}e^{\frac{(\pi+\delta -\pi \beta)}{\beta}t},
\nonumber \\
u(t)=\frac{\frac{(\delta+\pi)(1-\beta)}{\beta} u_0[\frac{k_0 }{c_0
z_0^{\frac{\beta-\sigma}{\sigma}}}-
F(t)]}{[(\frac{(\delta+\pi)(1-\beta)}{\beta}  +\delta u_0)\frac{k_0
}{c_0 z_0^{\frac{\beta-\sigma}{\sigma}} }-\delta u_0 G(t)]e^{
-\frac{(\delta+\pi)(1-\beta)}{\beta}t}-\delta u_0 [\frac{k_0 }{c_0
z_0^{\frac{\beta-\sigma}{\sigma}}}- F(t)]}\nonumber\\
 \lambda(t)=c_0^{-\sigma}z_0^{-\beta}e^{(\rho-\delta)t}z^{\beta},\nonumber\\
 \mu(t)= c_1 e^{(\rho-\delta )t}
\nonumber,\end{eqnarray}  where

\begin{eqnarray} \rho<\delta <\rho+\delta \sigma, \frac{\delta+\pi-\pi
\beta}{\beta}-\frac{\delta-\rho}{\sigma}
>0,\nonumber\\ F(t)=\int_0^t
z(t)^{\frac{\sigma-\beta}{\sigma}} e^{-(\frac{\delta+\pi-\pi
\beta}{\beta}-\frac{\delta-\rho}{\sigma} )t} dt, \nonumber\\
G(t)=\int_0^t z(t)^{\frac{\sigma-\beta}{\sigma}} e^{-\frac{\delta
\sigma-\delta+\rho}{\sigma} t} dt,
\label{sol2} \\
z(t)=
\frac{z^*z_0}{[(z^{*1-\beta}-z_0^{1-\beta})e^{-\frac{(1-\beta)(\delta
+\pi)}{\beta}t}+z_0^{1-\beta}]^{\frac{1}{1-\beta}} },\nonumber\\
 c_0
z_0^{\frac{\beta}{\sigma}}=\bigg(\frac{c_1
\delta}{(1-\beta)\gamma}\bigg)^{-\frac{1}{\sigma}}, \nonumber\\
\lim_{t\to\infty}
F(t)=\frac{k_0}{c_0z_0^{\frac{\beta-\sigma}{\sigma}}},\nonumber\\
\lim_{t\to\infty} \bigg[(\frac{(\delta+\pi)(1-\beta)}{\beta} +\delta
u_0)\frac{k_0 }{c_0 z_0^{\frac{\beta-\sigma}{\sigma}} }-\delta u_0
G(t) \bigg]=0,\nonumber\\ \lim_{t\to\infty} G(t)=
\frac{(\frac{(\delta+\pi)(1-\beta)}{\beta}  +\delta u_0)}{\delta
u_0}\lim_{t\to\infty} F(t), \nonumber\\z^*=\bigg(\frac{\beta
\gamma}{\delta +\pi} \bigg
)^{\frac{1}{\beta-1}}.\nonumber\end{eqnarray}
 Chilarescu
\cite{chil} derived same solution given on page 113 in Theorem 1 by
classical approach and utilized numerical simulations to evaluate
functions $F(t)$ and $G(t)$. Boucekkine and Ruiz-Tamarit
\cite{ruiz1} derived a similar solution and they expressed unknown
functions similar to $F(t)$ and $G(t)$ in terms of Hypergeometric
functions. Naz and Chaudhry \cite{naz2017} claimed that in
closed-form solutions (\ref{sol1}) and (\ref{sol2}) the expressions
for the variables $c(t)$, $k(t)$ are same but expressions for the
variables $h(t)$ and $u(t)$ are different. Thus closed-form solution
(\ref{sol1}) is different from closed-form solution (\ref{sol2}).

The uniqueness of solution discussed by Boucekkine and Ruiz-Tamarit
\cite{ruiz1}, Chilarescu \cite{chil} indicates that the expressions
for variables $h(t)$ and $u(t)$ in closed-form (\ref{sol1}) and
(\ref{sol2}) should be same. We equate expression for $h(t)$ and
$u(t)$ in (\ref{sol1}) and (\ref{sol2}), after simplifications, we
obtain following expression for unknown function $G(t)$ in terms of
$F(t)$:

\begin{eqnarray}
G(t)=F_*-(F_*-F(t))e^{\frac{(\delta+\pi)(1-\beta)}{\beta}
t}+\frac{\frac{(\delta+\pi)(1-\beta)}{\beta} F_*}{\delta u_0}
-\frac{e^{\frac{(\delta+\pi)(1-\beta)}{\beta} t}
\frac{(\delta+\pi)(1-\beta)}{\beta} }{\gamma
(1-\beta)[\rho-\delta(1-\sigma)]}\nonumber\\
\times \bigg[\sigma
z^{\beta-\frac{\beta}{\sigma}}e^{-(\frac{\delta+\pi(1-\beta)}{\beta}-\frac{\delta-\rho}{\sigma})
t}+\bigg(\gamma \beta (1-\sigma)-\big(\rho+\pi-
\pi\sigma\big)z(t)^{\beta-1}\bigg)(F_*-F(t))\bigg],\label{b1}
\end{eqnarray}
 provided following condition holds
\be \frac{\gamma(1-\beta)(\rho-\delta+\delta
\sigma)}{\delta}=\frac{u_0}{k_0}[\sigma c_0
z_0^{\beta-1}-(\rho+\pi-\pi \sigma)k_0 z_0^{\beta-1}+\beta
\gamma(1-\sigma)k_0], \label{con1} \ee  where $F_*=\frac{k_0}{c_0
z_0^{\frac{\beta-\sigma}{\sigma}}}$. It is important to mention here
that condition (\ref{con1}) arises systematically for the
closed-form solution (\ref{sol1}).

 In
(\ref{sol2}) the expression for $G(t)$ is \be G(t)= \int _0^t
z(s)^{\frac{\sigma-\beta}{\sigma}}e^{-\zeta s}ds,
\zeta=\bigg(\frac{\delta+\pi(1-\beta)}{\beta}-\frac{\delta-\rho}{\sigma}\bigg)-\frac{(\delta+\pi)(1-\beta)}{\beta}\label{b2}\ee

From (\ref{b1}) and  (\ref{b2}), we deduce that \begin{eqnarray}
 \int _0^t z(s)^{\frac{\sigma-\beta}{\sigma}}e^{-\zeta s}ds=
 F_*-(F_*-F(t))e^{\frac{(\delta+\pi)(1-\beta)}{\beta}
t}+\frac{\frac{(\delta+\pi)(1-\beta)}{\beta} F_*}{\delta u_0}
-\frac{e^{\frac{(\delta+\pi)(1-\beta)}{\beta} t}
\frac{(\delta+\pi)(1-\beta)}{\beta} }{\gamma
(1-\beta)[\rho-\delta(1-\sigma)]}\nonumber\\
\times \bigg[\sigma
z^{\beta-\frac{\beta}{\sigma}}e^{-(\frac{\delta+\pi(1-\beta)}{\beta}-\frac{\delta-\rho}{\sigma})
t}+\bigg(\gamma \beta (1-\sigma)-\big(\rho+\pi-
\pi\sigma\big)z(t)^{\beta-1}\bigg)(F_*-F(t))\bigg],\label{b3}\end{eqnarray}
provided  condition (\ref{con1}) holds.  If one can proof (\ref{b3})
as true only for that case the expressions for the variables $h(t)$
and $u(t)$ in closed-form (\ref{sol1}) and (\ref{sol2}) will be
same. Thus (\ref{sol1}) and (\ref{sol2})  provided by Naz and
Chaudhry \cite{naz2017} takes same form. This is consistent with
Chilarescu \cite{chil} and Boucekkine and Ruiz-Tamarit \cite{ruiz1}.

If $G(t)$ is different from (\ref{b3}) then multiple closed-form
solutions exist for the Lucas-Uzawa model for fairly general values
of parameters. It is an open question to prove (\ref{b3}) in
closed-form and not numerically.

\subsection{ Closed-form solution reported by  Naz et al \cite{naz2016} when $\sigma=\frac{\beta(\rho+\pi)}{2\pi\beta-\delta+\delta \beta-\pi}$}
 Naz et al \cite{naz2016} provided a closed-form solution
under a specific parametric restriction
$\sigma=\frac{\beta(\rho+\pi)}{2\pi\beta-\delta+\delta \beta-\pi}$
provided $2\pi\beta-\delta+\delta \beta-\pi>0$ to ensure that
$\sigma>0$. The parametric restriction arises automatically and it
was important to mention this solution which at the moment seems
purely mathematical solution. It might be interesting for economists
to test it empirically and it is an open question to test this
empirically.
\section{Closed-form solutions for Lucas-Uzawa model  with logarithmic utility preferences: Unique or multiple }
Chaudhry and Naz \cite{naz2018} provided two sets of closed-form
solutions. The first set of closed-form solutions for all variables
is\begin{eqnarray}  c(t)=c_0 z_0^{\beta}
e^{(\delta-\rho)t}z^{-\beta} , \nonumber\\
k(t)=c_0z_0^{\beta}
z(t)^{-1}e^{\frac{\delta}{\beta}t}\bigg(\frac{k_0
z_0^{1-\beta}}{c_0 }-F(t)\bigg),\nonumber\\
h(t)=\frac{\rho c_0 h_0}{c_0-\rho k_0}\bigg[\frac{1}{\rho
}e^{-(\rho-\delta)t}-
z(t)^{\beta-1}e^{\frac{\delta}{\beta}t}\bigg(\frac{k_0
z_0^{1-\beta}}{c_0 } - F(t) \bigg)\bigg],\nonumber\\
 u(t)=\frac{u_0 z_0^{\beta-1}(c_0-\rho k_0)\bigg(\frac{k_0
z_0^{1-\beta}}{c_0 }- F(t)\bigg)}{k_0\bigg[
e^{(\delta-\rho-\frac{\delta}{\beta})t}- \rho
z(t)^{\beta-1}\bigg(\frac{k_0 z_0^{1-\beta}}{c_0 } -F(t)\bigg)\bigg]},\nonumber\\
\mu(t)= \frac{A(1-\beta)}{\delta z_0^{\beta}c_0} e^{(\rho-\delta )t},\label{sol3}\\
\lambda(t)=\frac{1}{c_0z_0^{\beta}}e^{(\rho-\delta
)t}z^{\beta},\nonumber\end{eqnarray} where
\begin{eqnarray} F(t)= \int_0^t z(t)^{1-\beta}
e^{(\delta-\rho-\frac{\delta}{\beta})t} dt,\; \lim_{t\to\infty}
 F(t)=\frac{k_0
z_0^{1-\beta}}{c_0 },\nonumber\\
\frac{\delta u_0
z_0^{\beta}}{A(1-\beta)k_0 z_0}=\frac{\rho  }{c_0-\rho k_0},\nonumber\\\
z(t)=\frac{z^*
z_0}{\bigg(z_0^{1-\beta}+(z^{*1-\beta}-z_0^{1-\beta})e^{-\frac{(1-\beta)\delta}{\beta}t}\bigg)^{\frac{1}{1-\beta}}},\;z^*=\big(\frac{\beta
A}{\delta } \big )^{\frac{1}{\beta-1}}.\nonumber\end{eqnarray} The
second set of closed-form solutions for all variables as follows:
\begin{eqnarray}  c(t)=c_0 z_0^{\beta}
e^{(\delta-\rho)t}z^{-\beta} , \nonumber\\
k(t)=c_0z_0^{\beta}
z(t)^{-1}e^{\frac{\delta}{\beta}t}\bigg(\frac{k_0
z_0^{1-\beta}}{c_0 }- F(t)\bigg),\nonumber\\
 h(t)=\bigg[\bigg((\frac{\delta}{\beta}-\delta +\delta
u_0)\frac{k_0 z_0^{1-\beta}}{c_0 }-\delta u_0 G(t)
\bigg)e^{(\delta-\frac{\delta}{\beta})t}\nonumber\\ -\delta u_0
\bigg(\frac{k_0 z_0^{1-\beta}}{c_0
}-F(t)\bigg)\bigg]\frac{h_0c_0}{k_0z_0^{1-\beta}(\frac{\delta}{\beta}-\delta)}e^{\frac{\delta}{\beta}t},\nonumber\\
 u(t)=\frac{(\frac{\delta}{\beta}-\delta) u_0[\frac{k_0
z_0^{1-\beta}}{c_0 }- F(t)]}{[(\frac{\delta}{\beta}-\delta +\delta
u_0)\frac{k_0 z_0^{1-\beta}}{c_0 }-\delta u_0
G(t)]e^{(\delta-\frac{\delta}{\beta})t}-\delta u_0 [\frac{k_0
z_0^{1-\beta}}{c_0 }- F(t)]},\nonumber\\
\mu(t)= \frac{A(1-\beta)}{\delta z_0^{\beta}c_0} e^{(\rho-\delta )t},\label{sol4}\\
\lambda(t)=\frac{1}{c_0z_0^{\beta}}e^{(\rho-\delta
)t}z^{\beta},\nonumber\end{eqnarray} where
\begin{eqnarray} F(t)= \int_0^t z(t)^{1-\beta}
e^{(\delta-\rho-\frac{\delta}{\beta})t} dt,\;\lim_{t\to\infty}
 F(t)=\frac{k_0
z_0^{1-\beta}}{c_0 }, \nonumber\\
G(t)=\int_0^t z(t)^{1-\beta} e^{-\rho t} dt,\;\lim_{t\to\infty}
G(t)=\frac{(\frac{\delta}{\beta}-\delta+\delta u_0)}{\delta u_0
}\frac{k_0
z_0^{1-\beta}}{c_0 },\nonumber\\\
 z(t)=\frac{z^*
z_0}{\bigg(z_0^{1-\beta}+(z^{*1-\beta}-z_0^{1-\beta})e^{-\frac{(1-\beta)\delta}{\beta}t}\bigg)^{\frac{1}{1-\beta}}},\;z^*=\big(\frac{\beta
A}{\delta } \big )^{\frac{1}{\beta-1}}.\nonumber\end{eqnarray} It is
not difficult to show that closed-form solution {(\ref {sol4})}
which was derived by utilizing $I_1$ is exactly the same as the
solution found by Chilarescu and Sipos \cite{log2}. Chaudhry and Naz
\cite{naz2018} claimed that in closed-form solutions (\ref{sol3})
and (\ref{sol4}) the expressions for the variables $c(t)$, $k(t)$
are same but expressions for the variables $h(t)$ and $u(t)$ are
different. Thus closed-form solution (\ref{sol3}) is different from
closed-form solution (\ref{sol4}).

The uniqueness of solution discussed by Chilarescu  and Sipos
\cite{log2} indicates that the expressions for variables $h(t)$ and
$u(t)$ in closed-form (\ref{sol3}) and (\ref{sol4}) should be same.
We equate expression for $h(t)$ and $u(t)$ in (\ref{sol3}) and
(\ref{sol4}), after simplifications, we obtain following expression
for unknown function $G(t)$ in terms of $F(t)$:

\begin{eqnarray}
G(t)=F_*-(F_*-F(t))e^{(\frac{\delta}{\beta}-\delta)
t}+\frac{(\frac{\delta}{\beta}-\delta) F_*}{\delta u_0}
\nonumber\\-\frac{e^{(\frac{\delta}{\beta}-\delta) t}
(\frac{\delta}{\beta}-\delta) }{A (1-\beta)\rho}
\bigg[e^{(\delta-\rho-\frac{\delta}{\beta}) t}-\rho
z^{\beta-1}(F_*-F(t))\bigg],\label{bb1}
\end{eqnarray}
 provided following condition holds
\be\frac{\delta u_0 z_0^{\beta}}{A(1-\beta)k_0 z_0}=\frac{\rho
}{c_0-\rho k_0}, \label{conn1} \ee  where $F_*=\frac{k_0
z_0^{1-\beta}}{c_0 }$. It is important to mention here that
condition (\ref{conn1}) arises for the closed-form solution
(\ref{sol3}).

 In
(\ref{sol4}) the expression for $G(t)$ is \be G(t)=\int_0^t
z(t)^{1-\beta} e^{-\rho t} dt \label{bb2}\ee

From (\ref{bb1}) and  (\ref{bb2}), we deduce that \begin{eqnarray}
 \int_0^t
z(t)^{1-\beta} e^{-\rho t}
dt=F_*-(F_*-F(t))e^{(\frac{\delta}{\beta}-\delta)
t}+\frac{(\frac{\delta}{\beta}-\delta) F_*}{\delta u_0}\nonumber\\
-\frac{e^{(\frac{\delta}{\beta}-\delta) t}
(\frac{\delta}{\beta}-\delta) }{A (1-\beta)\rho} \
\bigg[e^{(\delta-\rho-\frac{\delta}{\beta}) t}-\rho
z^{\beta-1}(F_*-F(t))\bigg],\label{bb3}\end{eqnarray} provided
condition (\ref{conn1}) holds.  If one can proof (\ref{bb3}) as true
only for that case the expressions for the variables $h(t)$ and
$u(t)$ in closed-form (\ref{sol3}) and (\ref{sol4}) will be same.
Thus (\ref{sol3}) and (\ref{sol4}) provided by  Chaudhry and
\cite{naz2018} takes same form. This is consistent with Chilarescu
and Sipos \cite{log2}.

If $G(t)$ is different from (\ref{bb3}) then multiple closed-form
solutions exist for the Lucas-Uzawa model for fairly general values
of parameters. It is an open question to prove (\ref{bb3}) in
closed-form and not numerically.

\section{Conclusions}
 Naz and Chaudhry \cite{naz2017} established multiple closed-form
 solutions for the basic Lucas-Uzawa model. According to Boucekkine and
Ruiz-Tamarit \cite{ruiz1} and  Chilarescu \cite{chil} unique
closed-form solutions exist for the basic Lucas-Uzawa model. We
equated expressions for variables $h(t)$ and $u(t)$.  We provide
here condition for the unique closed-form solution. A similar
analysis  was carried out for the Lucas-Uzawa model with logarithmic
utility preferences. We propose open questions to prove (\ref{b3})
and (\ref{bb3}) in closed-form and not numerically. Can one test
empirically the closed-form solution reported by Naz et al
\cite{naz2016} when
$\sigma=\frac{\beta(\rho+\pi)}{2\pi\beta-\delta+\delta \beta-\pi}$.

 %We agree with claim of Boucekkine and
%Ruiz-Tamarit \cite{ruiz1} and  Chilarescu \cite{chil} that unique
%closed-form solution exists for the basic Lucas-Uzawa model. The
%simplest form of closed-form solution is provided by  Naz and
%Chaudhry \cite{naz2017}, it was possible due to systematic approach
%"partial Hamiltonian approach" based on Lie group theory.
 

\begin{thebibliography}{00}
 \bibitem{ruiz1}Boucekkine, R., \& Ruiz-Tamarit, J. R. (2008). Special functions for the study of economic dynamics: The case of the Lucas-Uzawa model.
       Journal of Mathematical Economics, 44(1), 33-54.
 \bibitem{chil}Chilarescu, C. (2011). On the existence and uniqueness of solution to the Lucas–Uzawa model. Economic Modelling, 28(1), 109-117.
       \bibitem{naz2017}Naz, R., \& Chaudhry, A. (2017). Comparison of Closed-Form Solutions for the Lucas-Uzawa Model via the Partial Hamiltonian Approach and the Classical Approach. Mathematical Modelling and Analysis, 22(4), 464-483.
  \bibitem{naz2016}Naz, R., Chaudhry, A., \& Mahomed, F. M. (2016). Closed-form solutions for the Lucas–Uzawa model of economic
  growth via the partial Hamiltonian approach. Communications in Nonlinear Science and Numerical Simulation, 30(1), 299-306.
  \bibitem{log1}Bethmann, D. (2013). Solving macroeconomic models with homogeneous technology and logarithmic preferences. Australian Economic Papers, 52(1), 1-18.
 \bibitem{log2}Chilarescu, C., \& Sipos, C. (2014). Solving Macroeconomic Models with Homogenous Technology and Logarithmic Preferences-A Note. Economics Bulletin, 34(1), 541-550.
 \bibitem{naz2018}Chaudhry, A., \& Naz, R. (2018). Closed-form solutions for the Lucas-Uzawa growth model with logarithmic utility preferences via the partial Hamiltonian approach. Discrete \& Continuous Dynamical Systems-Series S, 11(4).
         \end{thebibliography}
\end{document}